\documentclass[a4paper]{article}
\pdfoutput=1

\usepackage[english]{babel}
\usepackage{icrc2013}

\usepackage{pstricks,pst-grad}

\title{Using Raster Scans of Bright Stars to Measure the Relative Total Throughputs of Cherenkov Telescopes}

\shorttitle{Cherenkov Telescope Raster Scans}

\authors{
Sean Griffin$^{1}$,
and David Hanna$^{1}$
}

\afiliations{
$^1$ Department of Physics \\
McGill University \\
Montreal, QC H3A 2T8, Canada}

\email{griffins@physics.mcgill.ca}

\abstract{Gamma-ray astronomy at energies in excess of 100 GeV is carried out using arrays of imaging Cherenkov telescopes. Each telescope comprises a large reflector, of order 10 m diameter, made of many mirror facets, and a camera consisting of a matrix of photomultiplier pixels. Differences in the total throughput between nominally identical telescopes, due to aging of the mirrors and PMTs and other effects, should be monitored to reduce possible systematic errors. One way to directly measure the throughput of such telescopes is to track bright stars and measure the photocurrents produced by their light falling on camera pixels. We have developed such a procedure using the four telescopes in the VERITAS array. We note the technique is general, however, and could be applied to other imaging Cherenkov experiments. For this measurement, a raster scan is performed on a single star such that its image is swept across the central pixels in the camera, thus providing a statistically robust set of measurements in a short period of time to reduce time-dependent effects on the throughput. Photocurrents are measured using the starlight-induced baseline fluctuations of the pixel outputs, as recorded by the standard readout electronics. In this contribution we describe details of the procedure and report on feasibility studies carried out during the 2012-2013 observing season.}

\keywords{Cherenkov Telescopes, Calibration}

\begin{document}
\maketitle

\section{Introduction}

Very-high-energy (VHE) gamma-ray astronomy makes use of arrays of imaging
atmospheric Cherenkov telescopes (IACTs). 
Observations made with more than one telescope achieve better background 
rejection, improved energy and angular resolution, and are immune to the 
effects of local muons.
An issue that arises when using multiple telescopes is that of 
their relative calibration.
Differences between nominally identical telescopes can come about from 
differential aging of mirror facets and photomultiplier tubes (PMTs) 
or from different maintenance or upgrade schedules.
A simple parameter that can be used to correct for the overall effect of such 
changes is the relative total throughput of a telescope.
This parameter can be estimated using a variety of techniques, such as the 
inclusive rate for cosmic-ray showers~\cite{lebohec}, 
analysis of shower-image sizes~\cite{hofmann},  
signals from local muons~\cite{vacanti} acquired using
special triggers, and observations of scattered light from a distant
laser beam~\cite{shepherd, hui}. 
A solid understanding of a telescope's calibration will result in the same 
number emerging from each of the techniques and the origin of changes to the
number can be determined by examining data from component-specific 
calibration procedures such as light-pulse PMT calibration~\cite{flasher}
and whole-dish mirror reflectivity measurements~\cite{reflectivity}.

In this contribution we describe a total-throughput measurement procedure based
on using photocurrents induced by the image of a bright star falling on PMTs
in a telecope's camera. 
The method was developed for the VERITAS array and we report here on initial 
tests made with that instrument. However the method is quite 
general and can be used 
for other arrays.
Initial tests were conducted using magnitude 7 stars with spectra very different
from the standard Cherenkov spectrum relevant to air-shower detection.
We are currently exploring the use of ultraviolet filters, already 
acquired for observing under 
bright moonlight, to extend this technique to shorter wavelengths.

\section{The VERITAS IACT Array}

VERITAS comprises an array of four IACTs located at the Whipple Observatory 
at the base of Mount Hopkins in southern Arizona~\cite{holder,weekes}. 
Each of the telescopes is based on a 12-m diameter Davies-Cotton reflector 
focussing light onto a 499-pixel camera made from close-packed 
Hamamatsu R10560 PMTs coupled to conical light concentrators.
Each reflector is made up of 345 identical mirror facets, the alignment of which
is such that the on-axis point-spread-function is smaller than a pixel 
diameter~\cite{mccann}.

\section{Raster Scanning}
The total throughput of one of the telescopes could be measured 
by tracking a single star and measuring the photocurrent from the camera's
central pixel.
However it is statistically more powerful to illuminate several pixels, 
in sequence, in order to average out effects such as differences in 
the wavelength dependence of quantum efficiency in different PMTs.
The standard tracking software for VERITAS is designed to map a 
telescope's nominal pointing direction onto the central pixel and to change that
to an arbitrary pixel required modifications. 
Such modifications were implemented as part of the VERITAS mirror alignment 
scheme~\cite{mccann} whereby a raster scan over a grid centred on a bright star
is performed. 
We have adapted this scanning technique to cause the image of a star to 
sweep over the central pixels of the camera in a controlled fashion and with 
a grid step that is small enough to obtain data with the star image very 
close to the centre of a given pixel. 

\begin{figure}[h]
  \centering
  \includegraphics[width=0.5\textwidth,clip=true,trim=0 5.75cm 0 5cm]{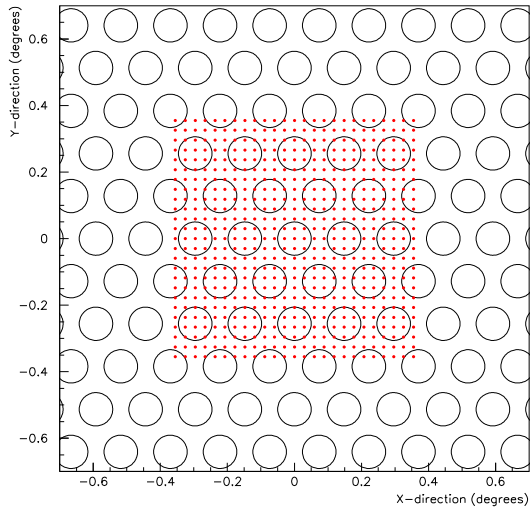}
  \caption{Diagram of the central region of a VERITAS camera showing the 
positions of the PMTs and the grid of points where the centroid of a star 
image is expected to be during a raster scan.
}
  \label{camera}
 \end{figure}

For the data reported on here we used a 25-by-25 array of pointings with a 
step size of 0.03 degrees.
A VERITAS pixel has a field-of-view diameter of 0.15 degree so the grid 
should cover a square array of order 25 pixels.
The mis-match between the hexagonal nature of the PMT positions and the square
scan grid reduces this to 23 pixels (Figure~\ref{camera}).

\section{Photocurrent Measurement}

\begin{figure}[h]
  \begin{center}
  \includegraphics[width=0.5\textwidth,clip=true,trim=0 5.75cm 1.5cm 5cm]{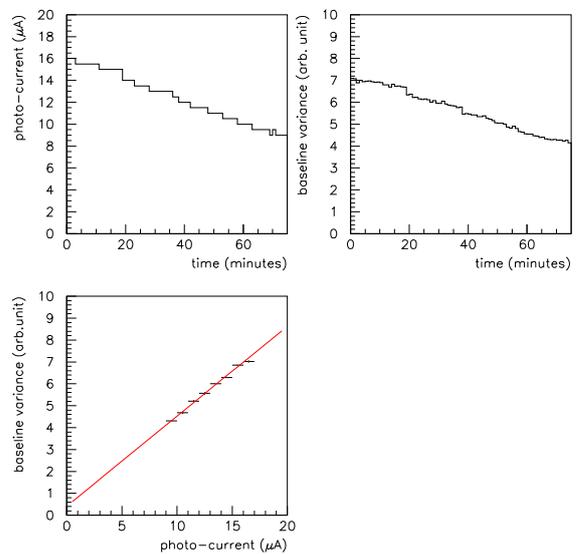}
  \caption{Correlation of FADC baseline variance with PMT photocurrent for a 
single pixel.
In the upper left panel the photocurrent is plotted as a function 
of time. The corresponding baseline variance is plotted in the 
upper right panel.
The average variance is plotted as a function of photocurrent
in the lower left panel, together with a linear fit.}
  \label{v-vs-i}
  \end{center}
 \end{figure}

The readout electronics for a VERITAS pixel consist of a preamp in the PMT base
followed by an amplifier and a 500 MS/s FADC located in the electronics shed 
under the telescope. 
The signals are AC coupled; a capacitor before the preamp blocks the DC 
photocurrent. However, a resitive path to ground upstream of the capacitor  is
provided for purposes of monitoring this current.
Low-resolution (0.5~$\mu$A step size) measurements are available, mainly to 
allow for switching off the high-voltage to a PMT in case of excess currents.
For our purposes we need finer granularity.
To obtain this we make use of the fact that baseline fluctuations 
(pedestal variations) as recorded by the FADC readout can be used as a proxy 
for photocurrent. 
This can be motivated by a simple model that posits current as coming from
a stream of single-photoelectron 
pulses approximated as narrow digital pulses. 
The empirical proof of the correlation is 
shown in Figure~\ref{v-vs-i}.

Data for this figure were extracted from standard observing runs made
under partial moonlight but while the moon was setting so the currents vary 
over an interesting range. In the upper left panel we plot the current readings for an arbitrary pixel as a
function of time in minutes. 

In the upper right panel we plot the corresponding smoothed baseline variances. 
For every event (approximately 300 times per second for these runs; the rate is random and dominated by cosmic ray triggers) 
a 16-sample FADC trace is recorded for each pixel and the variance can be 
calculated. 
Most traces are empty except for fluctuations due to night sky background photons (the scale of these fluctuations is much 
less than the scale of Cherenkov pulses). However, since the data were acquired under normal trigger 
conditions, there are some 
traces with Cherenkov pulses in them and these cause long
tails in the variance distributions.
To deal with this we use a ``mean of medians'' technique.
First, we divide the data stream into groups of 300 events. 
These are further subdivided into 60 subgroups of 5 and we average the 
60 medians
from each 5-member subgroup, obtaining an average
variance estimate approximately once per second. 

In the lower left panel we plot the average variances as a function of current.
The linear correlation is evident and justifies the use of baseline variance 
as a proxy for current in the following.
The slopes of the fitted line from plots like that in Figure~\ref{v-vs-i} are 
used to convert the increment in baseline variance due to the effect of the star
to an increment in photocurrent.
 
\section{Test Results}

Raster-scan data for this study were acquired on two separate occasions. 
For each run a magnitude 7 star was chosen as the target and the telescopes
were slewed to its coordinates.
The raster scan was then performed and at the end the telescopes returned to 
their nominal tracking directions.
The scan was carried out with a one-second dwell time at each pointing and one 
second between points for slewing and settling so the 625-point scan took 
just over 20 minutes.
The target star was selected to be rising and close to transit; even though the
run was reasonably short, we wanted to avoid systematic effects due to 
changes in flux due to atmospheric absorption.
Data were acquired by externally triggering the array at a fixed rate of 300 Hz.

\begin{figure*}[htb]
  \centering
  \includegraphics[width=1.0\textwidth,clip=true,trim=0 14.75cm 0 5cm]{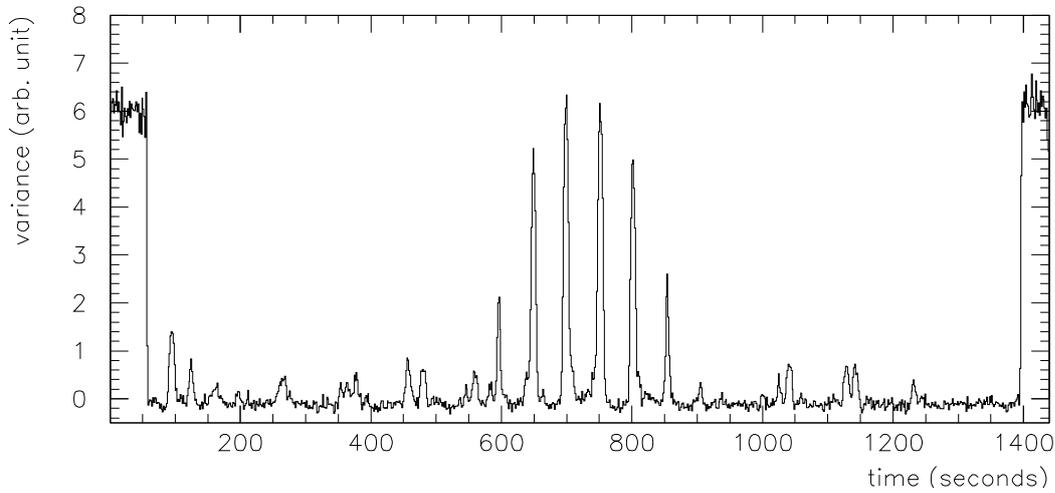}
  \caption{Background-subtracted 
variances vs. time for the central pixel of a 
VERITAS telescope during a raster scan run. 
The run begins with the telescope tracking a star, causing increased
baseline fluctuations in the central pixel. 
As the scan continues, the star image leaves the central pixel and does not return 
until the middle of the run where it is seen causing different increases in 
baseline fluctations, depending on its overlap with the pixel, as it is swept 
back and forth 
across the camera. At the end of the run the telescope returns to nominal 
tracking and the fluctuations increase again. Nearby stars with lower 
brightness are responsible for the activity elsewhere in the plot.
}
  \label{raster-0}
 \end{figure*}

\begin{figure*}[htb]
  \centering
  \includegraphics[width=1.0\textwidth,clip=true,trim=0 14.75cm 0 5cm]{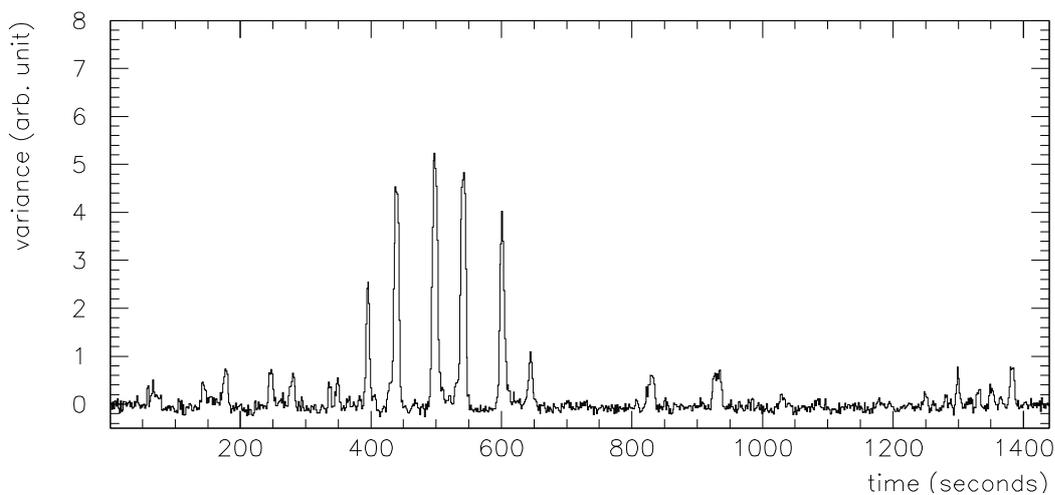}
  \caption{As in the previous figure but for a different pixel. There is no
activity at the beginning and end of the run since that is unique to the 
central pixel. Regions of increased variance are shifted in 
accordance with the pixel's location in the camera.
}
  \label{raster-6}
 \end{figure*}

Results from the central pixel in one of the telescopes are shown in 
Figure~\ref{raster-0} where we plot baseline variance (current proxy)
as a function of time in seconds.
The elevated currents at the beginning and end of the run are due to the 
tracking of the target star such that its image is contained within the 
central pixel. 
This feature is absent in Figure~\ref{raster-6} where data from an off-centre
pixel are plotted.
In both figures one sees structure resulting from the scan where currents rise
and fall as the star image is swept across the pixel field-of-view and the 
maximum of each peak rises and falls as the distance of the scan line from 
the centre of the pixel varies. In the following, we use the amplitude of the 
largest peak to make an estimate of 
the current that would result if the star's image were exactly centred on 
the pixel.  

\begin{figure}[h]
  \centering
\includegraphics[width=0.5\textwidth,clip=true,trim=0 5.75cm 0 5cm]{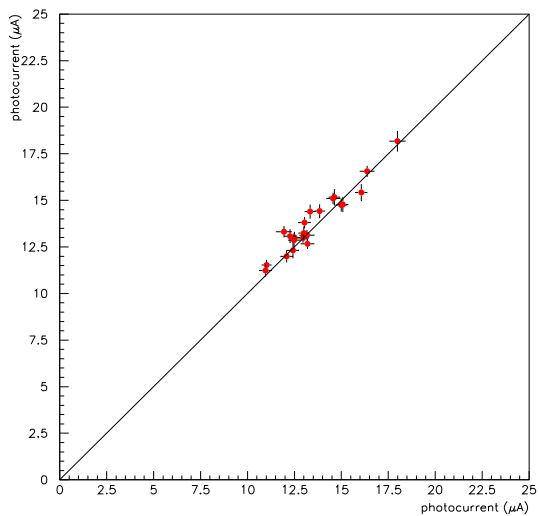}

  \caption{Reproducibility of raster-scan data. Peak currents from one 
scan are plotted against the corresponding currents from a scan performed 
on the previous night. 
Each point corresponds to a different pixel.
}
  \label{raster-repro}
 \end{figure}

In Figure \ref{raster-repro} we plot the peak currents achieved on one night
vs the peak currents from the previous night for a single telescope.
It is clear that the results obtained are reproducible over the short term and 
that the statistical errors are understood.

\begin{figure}[!h]
  \centering
  \includegraphics[width=0.5\textwidth,clip=true,trim=0 5cm 0 5cm]{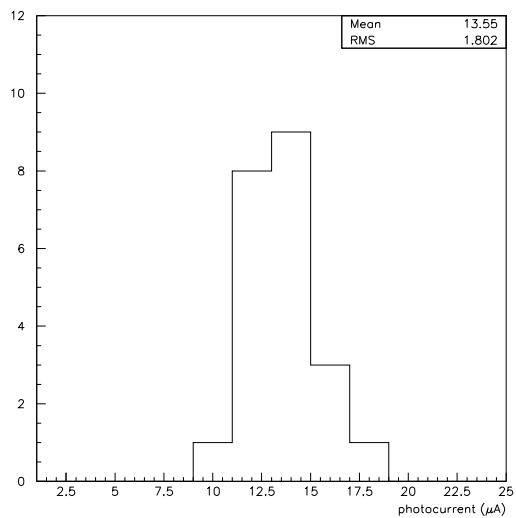}
  \caption{Peak currents from pixels in a VERITAS telescope for one of the runs used in Figure~\ref{raster-repro}.
}

  \label{raster-dat}
 \end{figure}

The peak currents from this telescope for a single night ({\it i.e.} the x 
projection of data plotted in Figure~\ref{raster-repro}) are plotted in 
Figure~\ref{raster-dat}.
The dispersion is relatively large; the RMS is 
slightly more than 10\% of the mean. 
The reasons for this are under study but may be partly due 
to the fact that the star's image does not cross over the 
exact center of every camera pixel.

Similar results have been obtained from all telescopes in the VERITAS array
and we are currently evaluating the level of systematic errors to be expected.
Already the statistical errors indicate that we can expect to measure  
differences in relative throughput of a few percent.

\section{Conclusions}

We have tested a method for measuring the net throughputs of different 
telescopes in an array. 
The procedure requires no specialized equipment and can be carried out in 
less than 30 minutes, possibly during periods where moonlight or non-optimal 
weather lessen the competition for observing time. 
The initial results are very encouraging and we expect to pursue this in the 
future to look at long-term stability and possible improvements to the 
precision of the method.

\vspace*{0.5cm}
\normalsize{{\bf Acknowledgments:}{
We warmly thank our colleagues in the VERITAS collaboration for their support
of this work and for assistance with data acquisition.
VERITAS research is supported by grants from the U.S. Department of Energy Office of Science, the U.S. National Science Foundation and the Smithsonian Institution, by NSERC in Canada, by Science Foundation Ireland (SFI 10/RFP/AST2748) and by STFC in the U.K. We acknowledge the excellent work of the technical support staff at the Fred Lawrence Whipple Observatory and at the collaborating institutions in the construction and operation of the instrument. }}

\end{document}